# Electronically Tunable Voltage-Mode Biquad Filter/Oscillator Based On CCCCTAs

Sajai Vir Singh

Department of Electronics and Communication Engineering,
Jaypee Institute of Information Technology, Sect-128, Noida-201304, India
sajaivir75@gmail.com

Gungan Gupta

Department of Electronics and Communication Engineering,
RKGIT, Ghaziabad -201001, India
Gunjan.techno@gmail.com

Rahul Chhabra

Department of Electronics and Communication Engineering,
Jaypee Institute of Information Technology, Sect-128, Noida-201304, India
rahul.december90@gmail.com

Kanika Nagpal

Department of Electronics and Communication Engineering,
Jaypee Institute of Information Technology, Sect-128, Noida-201304, India
kanika.nagpal21@gmal.com

Devansh

Department of Electronics and Communication Engineering,
Jaypee Institute of Information Technology, Sect-128, Noida-201304, India
kumar_devansh@yahoo.co.in

*Abstract*— **In this paper, a circuit employing current controlled current conveyor trans-conductance amplifiers (CCCCTAs) as active element is proposed which can function both as biquad filter and oscillator. It uses two CCCCTAs and two capacitors. As a biquad filter it can realizes all the standard filtering functions (low pass, band pass, high pass, band reject and all pass) in voltage-mode and provides the feature of electronically and orthogonal control of pole frequency and quality factor through biasing current(s) of CCCCTAs. The proposed circuit can also be worked as oscillator without changing the circuit topology. Without any resistors and using capacitors, the proposed circuit is suitable for IC fabrication. The validity of proposed filter is verified through PSPICE simulations.**

I. INTRODUCTION

In analogue signal processing applications such as communication system, instrumentation and control engineering, oscillators and filters are frequently used as two analog building blocks. An oscillator is used in transmitters to create carrier waves, waveforms created for the purpose of transmitting information. They are also used in radios as a way of changing the modulation of information-carrying waveforms to allow the device (the radio receiver) to receive and interpret the information carrying waveforms [1]. Analog filters find many applications in video signal enhancement, graphic equalizer in hi-fi systems, dual tone multi-frequency (DTMF) for use in touch-tone dialing in the telephone market, phase locked loop and cross over network used in three way high fidelity loud speaker [2]. So in recent past, there has been greater emphasis on design of universal biquad active filters and oscillators and hence, several voltage-mode filters and oscillators using different current-mode active elements are proposed in the literatures [3-21]. However, from our investigations, there are seen that the voltage-mode oscillators and filters reported in the previous literatures [3-18] require too many components. In addition, each circuit can work only one function, either universal biquad filter [3-11] or oscillator [4-18]. Very few voltage-mode circuits are available in the literatures [19-21] which can be used as both filters and oscillators. The circuit [19] uses three DVCCs, two capacitors, three resistors while the circuit [20] uses two CCCDBAs, two capacitors. Moreover, another circuit [21] employs single DBTA and four passive elements. Each circuit [19-21] realizes all the five standard filtering functions as biquad filter and sinusoidal quadrature oscillations as oscillator. However, in all above three circuits [19-21], oscillator structure is obtained with slight modification in the filter structure i.e. both filter and oscillator function can't be obtained with out modification in circuit topology. In addition, two of the circuits [19, 21] do not provide the feature of electronic tunability of pole frequency independent of quality factor.

In this paper, a new electronically tunable circuit topology employing two CCCCTAs and two capacitors is proposed. This topology can realize three-input single output voltage-mode biquad filter and oscillator with out changing the circuit configuration. As a biquad filter it can realizes all the standard filtering functions (low pass, band pass, high pass, band reject and all pass) in voltage-mode and provides the feature of electronic tunability of pole frequency independent of quality factor through biasing current(s) of CCCCTAs. As oscillator, circuit provides three voltage-mode sinusoidal oscillations. The workability of proposed filter is verified through PSPICE, the industry standard tool.

## II. CCCCTA Description

Current controlled current conveyor trans-conductance amplifier (CCCCTA) has received considerable attention as current-mode active element since last few years [22]. CCCCTA is a combination of a CCCII followed by an OTA. The main advantage of CCCCTA is its electronic tuning ability through the parasitic resistance at terminal X and trans-conductance parameter ($g_m$), hence it does not need a resistor in practical applications. Subsequently, the CCCCTA based circuits realizations occupy less chip area. This device can be operated in both current as well as voltage-modes, providing flexibility to the circuit designers. In addition, it can offer several advantages such as high slew rate, wider bandwidth and simpler implementation, associated with current-mode active elements. All these advantages together with its current-mode operation make the CCCCTA, a promising building block for realizing active filters and oscillators [9,11]. The schematic symbol of CCCCTA is shown in Fig.1 where X and Y are input terminals which have low and high impedance level, respectively. It consists of one Z stage with high output impedance terminal(s). The current through the terminal Z follows the current through the X terminal. The voltage across the auxiliary Z terminal is transferred to a current at one or more trans-conductance output terminals (+O or –O or both type) by a trans-conductance parameter ($g_m$) which is electronically controllable by an external bias current ($I_S$). $R_X$ is the parasitic resistance at X terminal of the CCCCTA which depends upon the biasing currents $I_B$ of the CCCCTA. The CCCCTA properties can be described by the following equations

$$V_{Xi} = V_{Yi} + I_{Xi}R_{Xi} \ , \ I_{Zi} = I_{Xi} \ , \ I_{\pm O} = \pm g_{mi}V_{Zi} \quad (1)$$

where $R_{xi}$ and $g_{mi}$ are the parasitic resistance at x terminal and transconductance of the i$^{th}$ CCCCTA, respectively. $R_{xi}$ and $g_{mi}$ depend upon the biasing currents $I_{Bi}$ and $I_{Si}$ of the CCCCTA, respectively. For BJT model of CCCCTA [11], $R_{xi}$ and $g_{mi}$ can be expressed as

$$R_{Xi} = \frac{V_T}{2I_{Bi}} \quad \text{and} \quad g_{mi} = \frac{I_{Si}}{2V_T} \quad (2)$$

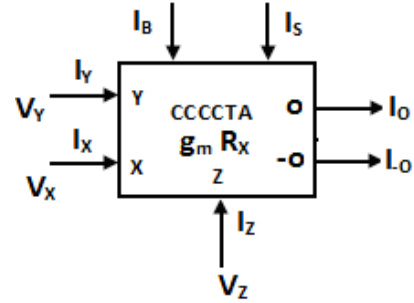

Figure1. CCCCTA Symbol

## III. Proposed Circuit

### A. The Proposed Circuit operating as Universal Voltage-Mode Biquad Filter

The proposed circuit operating as universal voltage-mode biquad filter is shown in Fig.2. It is based on two CCCCTAs and two capacitors. Routine analysis of the proposed biquad filter yields the following output voltage

$$V_0 = \frac{V_1 s^2 C_1 C_2 R_{X1} + V_3 s g_{m1} R_{X1} C_2 + V_2 s C_2 + V_2 g_{m1}}{s^2 C_1 C_2 R_{X1} + s(1 - g_{m2} R_{X1})C_2 + g_{m1}} \quad (3)$$

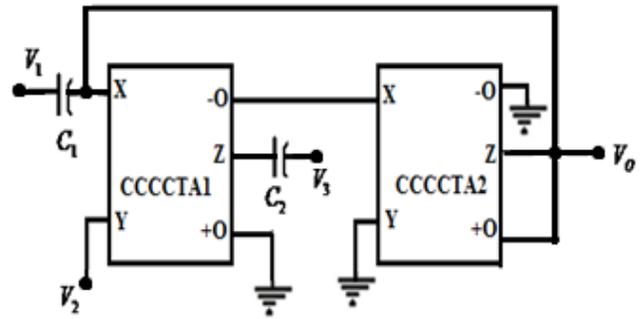

Figure 2. Proposed circuit working as voltage-mode universal biquad filter

It is clear from (3) that the proposed circuit can be used as three input single output voltage-mode biquad filter by maintaining $g_{m2}R_{X1} \ll 1$ and provides various filtering responses in voltage-mode through appropriate selection of input voltages which are as follows:

(i) High pass response, with $V_1 = 1$, $V_2 = V_3 = 0$

(ii) Band pass response, with $V_1 = V_2 = 0$, $V_3 = 1$

*(iii)* Low pass response, with $V_1 = 0$, $V_2 = 1$, $V_3 = -1$ and

$g_{m1} R_{X1} = 1$

*(iv)* Band reject response, with $V_1 = V_2 = 1$, $V_3 = -1$ and

$g_{m1} R_{X1} = 1$

*(v)* All pass response, with $V_1 = V_2 = 1$, $V_3 = -1$ and

$g_{m1} R_{X1} = 2$

Thus, the circuit is capable of realizing all the standard filtering responses in voltage mode from the same configuration. The pole frequency ($\omega_o$), quality factor (Q) and bandwidth (BW) $\omega_o/Q$ of each filter responses can be expressed as

$$\omega_o = \left(\frac{g_{m1}}{C_1 C_2 R_{X1}}\right)^{\frac{1}{2}}, Q = \frac{1}{(1-g_{m2}R_{X1})}\left(\frac{C_1 R_{X1} g_{m1}}{C_2}\right)^{\frac{1}{2}} \quad (5)$$

Substituting intrinsic resistances and transconductance values as depicted in (2) and $I_{S2} \ll I_{B1}$, it yields

$$\omega_o = \frac{1}{V_T}\left(\frac{I_{S1} I_{B1}}{C_1 C_2}\right)^{\frac{1}{2}}, \quad Q = \frac{1}{2}\left(\frac{I_{S1} C_1}{I_{B1} C_2}\right)^{\frac{1}{2}} \quad (6)$$

From (6), by maintaining the ratio $I_{B1}$ and $I_{S1}$ to be constant, it can be remarked that the pole frequency can be adjusted by $I_{B1}$ and $I_{S1}$ without affecting the quality factor. The active and passive sensitivities of the proposed biquad filter as shown in Fig.2, can be found as

$$S^{\omega_o}_{C_1,C_2} = -\frac{1}{2}, S^{\omega_o}_{I_{S1},I_{B1}} = \frac{1}{2}, S^{\omega_o}_{I_{S2},I_{B2}} = 0 \quad (7)$$

$$S^{Q}_{I_{B1},C_2} = -\frac{1}{2}, S^{Q}_{I_{S1},C_1} = \frac{1}{2}, S^{Q}_{I_{S2},I_{B2}} = 0 \quad (8)$$

From the above results, it can be observed that all the sensitivities are low and within half in magnitude.

### B. The Proposed Circuit Operating as Quadrature Oscillators

If no input voltage signal is applied in the circuit of Fig.2, a quadrature oscillator circuit is further realized. The resulting circuit working as oscillator is shown in Fig.3. The circuit analysis yields the following characteristic equation

$$s^2 C_1 C_2 R_{X1} + s(1 - g_{m2} R_{X1})C_2 + g_{m1} = 0 \quad (9)$$

Identify applicable sponsor/s here. *(sponsors)*

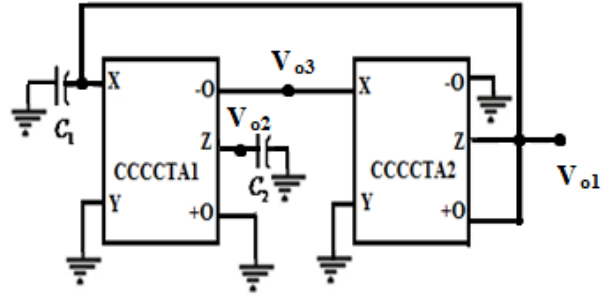

Figure 3. Proposed circuit working as oscillator

At the frequency of oscillation, with $s = j\omega$, the equation gives the frequency of oscillation (FO) and condition of oscillation (CO) as

$$\text{FO: } \omega_o = \left(\frac{g_{m1}}{C_1 C_2 R_{X1}}\right)^{\frac{1}{2}} = \frac{1}{V_T}\left(\frac{I_{S1} I_{B1}}{C_1 C_2}\right)^{\frac{1}{2}} \quad (10)$$

$$\text{And CO: } g_{m2} = \frac{1}{R_{X1}} \quad (11)$$

From (10), it can be seen that frequency of oscillation ($\omega_o$) can be controlled by biasing current $I_{S1}$ without affecting condition of oscillation. The condition of oscillation can also be adjusted by $g_{m2}$ (or $I_{S2}$) without affecting frequency of oscillation. Therefore, the frequency of oscillation and the condition of oscillation of the proposed quadrature oscillator circuit can be controlled electronically and independently. Furthermore, the quadrature sinusoidal signal outputs can be obtained at $V_{O1}$, $V_{O2}$ and $V_{O3}$.

### IV. SIMULATION RESULTS

To validate the theoretical analysis, the proposed circuit was simulated through PSPICE. In simulation, the CCCCTA was realized using BJT model as shown in Fig.4, with the transistor model of HFA3096 mixed transistors arrays [11] and was biased with ±1.75V DC power supplies. The SPICE model parameters are given in Table1. Firstly, the operation of the proposed circuit as voltage-mode biquad filter as shown in Fig. 2 was verified. The proposed biquad filter was designed for Q=1 and $f_o=\omega_o/2\pi=196.71$ KHz. The active and passive components were chosen as $I_{B1}=I_{B2}=80\mu A$, $I_{S1}=320\mu A$, $I_{S2}=2\mu A$ and $C_1=C_2=5nF$. Fig.5 shows the simulated voltage gain and phase responses of the LP, HP, BP, BR and AP. The simulation results show the simulated pole frequency as 184.77 KHz that agree quite well with the theoretical analysis. Fig.6 shows magnitude responses of BP function where $I_{B1}$ and $I_{S1}$ are equally set and changed for several values, by keeping its ratio to be constant for constant Q(=0.5). Other parameters were chosen as $I_{B2}=80\mu A$, $I_{S2}=2\mu A$, and $C_1=C_2=5nF$. The pole frequency (in Fig.6) is found to vary as 36KHz, 72KHz, 142KHz and 276KHz for four values of $I_{B2}=I_{S2}$ as 30μA, 60μA, 120μA and 240μA, respectively, which shows that pole frequency can be electronically adjusted without affecting the quality factor.

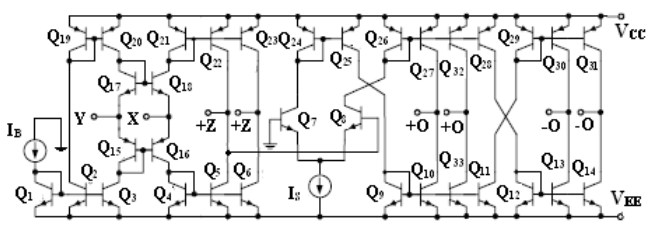

Figure 4. BJT implementation of CCCCTA

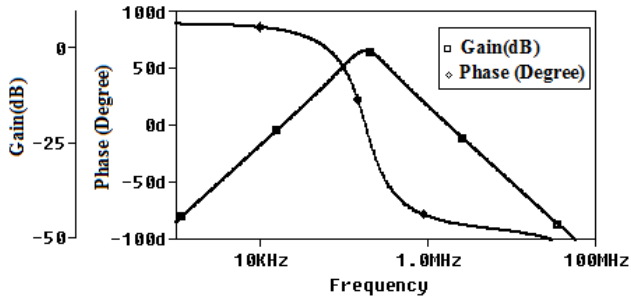

(a)

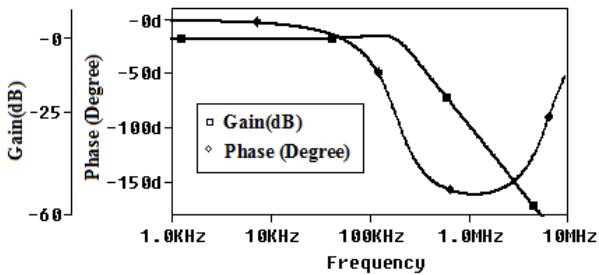

(b)

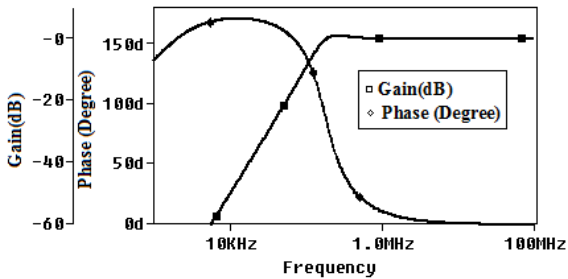

(c)

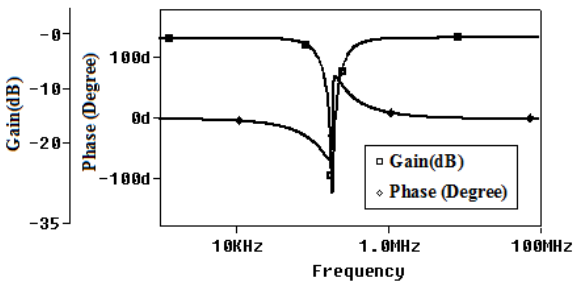

(d)

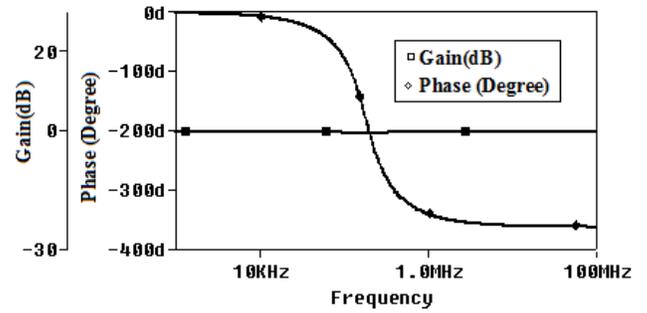

(e)

Figure 5. Voltage gain and phase responses of (a) BP (b) LP (c) HP (d) BR (e) AP for the proposed circuit as biquad filtering operation of Fig. 2

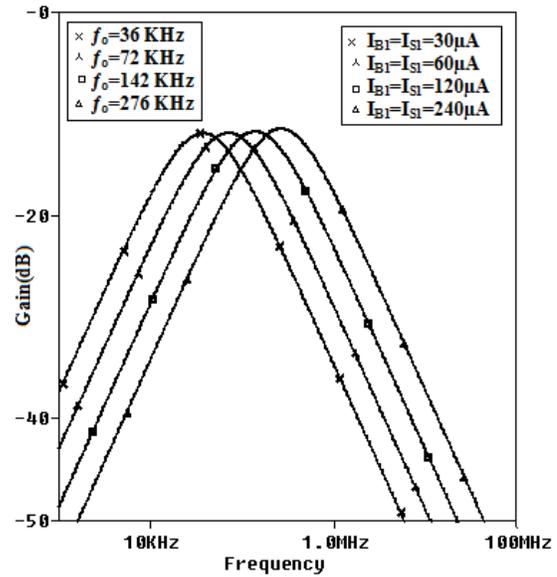

**Fig.6** Band pass responses of the proposed circuit as biquad filter for different value of $I_{B1}=I_{S1}$

Next, in order to confirm the above given theoretical analysis of the proposed circuit as oscillator in Fig.3, it was also simulated using PSPICE simulation. To obtain the sinusoidal oscillations with the oscillation frequency of 130 KHz, the active and passive components were chosen as $I_{B1}=56.5\mu A$, $I_{B2}=45\mu A$, $I_{S1}=200\mu A$, $I_{S2}=225\mu A$ and $C_1=C_2=5nF$. The simulated sinusoidal oscillations result is shown in Fig.7. The simulated oscillation frequency was measured as 128 KHz which is quite close to the theoretical value of 130 KHz.

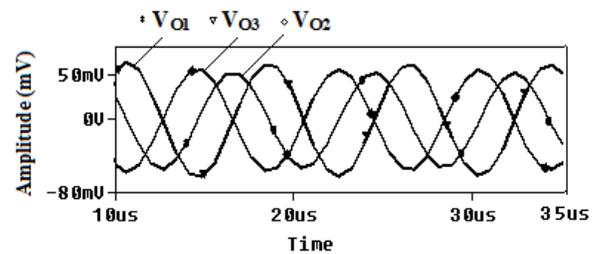

Figure 7. Quadrature outputs of circuit of Fig.3

## V. CONCLUSION

In this paper, a new circuit topology is proposed which can function both as biquad filter and oscillator with out change in circuit configuration. It uses only two CCCCTAs and two capacitors. As a biquad filter it realizes all the standard filtering functions in voltage-mode and provides the feature of electronic orthogonal control of pole frequency and quality factor through biasing current(s) of CCCCTAs. As oscillator the frequency of oscillation and the condition of oscillation of the proposed circuit can be controlled electronically and independently. The validity of proposed filter is verified through PSPICE simulations.

**Table1:** The SPICE model parameters of HFA3096 mixed transistors arrays

| | |
|---|---|
| .model npn | Is=1.80E-17, Xti=3.20, Eg=1.167, Vaf=151.0, Bf=1.10E+02, Ne=2.000, Ise=1.03E-16, IKf=1.18E-02, Xtb=2.15, Br=8.56E-02, IKr=1.18E-02, Rc=1.58E+02, Cjc=2.44E-14, Mjc=0.350, Vjc=0.633, Cje=5.27E-4, Mje=0.350, Vje=1.250, Tr=5.16E-08, Tf=2.01E-11, Itf=2.47E-02, Vtf=6.62, Xtf=25.98, Rb=8.11E+02, Ne=2, Isc=0, Fc=.5 |
| .model pnp | Is=8.40E-18, Xti=3.67, Eg=1.145, Vaf=57.0, Bf=9.55E+01, Ne=2.206, Ise=3.95E-16, IKf=2.21E-03, Xtb=1.82, Br=3.40E-01, IKr=2.21E-03, Rc=1.43E+02, Cjc=3.68E-14, Mjc=0.333, Vjc=0.700, Cje=4.20E-14, Mje=0.560, Vje=.8950, Tr=2.10E-08, Tf=6.98E-11, Itf=2.25E-02, Vtf=1.34, Xtf=12.31, Rb=5.06E+02, Ne=2, Isc=0, Fc=.5 |

## AUTHORS PROFILE

**Sajai Vir Singh** was born in Agra, India. He received his B.E. degree (1998) in Electronics and Telecommunication from NIT Silchar, Assam (India), M.E. degree (2002) from MNIT Jaipur, Rajasthan (India) and Ph.D. degree (2011) from Uttarakhand Technical University. He is currently working as Assistant Professor in the Department of Electronics and Communication Engineering of Jaypee Institute of Information Technology, Noida (India) and has been engaged in teaching and design of courses related to the design and synthesis of Analog and Digital Electronic Circuits. His research areas include Analog IC Circuits and Filter design. He has published more than 25 research papers in various International Journal/Conferences.

**Gunjan Gupta** received B.Tech degree (2006) in Electronics & Telecommunication and M.Tech degree in VLSI Design from U. P. Technical University, Lucknow, Uttar Pradesh, India. She has been with RKGIT, Ghaziabad affiliated to U. P. Technical University, Lucknow, Uttar Pradesh, India as an Assistant Professor for 6 years. and is currently pursuing her Ph.D from Jaypee University of Information Technology Waknaghat, India. Her research area is analog signal processing.

**Rahul Chhabra** was born in Dehradun, Uttrakhand, India. He is a 4[th] year student and pursuing a bachelor of technology (B.Tech) degree in Electronics and Communication Engineering from Jaypee Institute of Information Technology, Noida. His research interest is designing of analog circuit.

**Kanika Nagpal** was born in Delhi, India. She is a 4[th] year student and pursuing a bachelor of technology (B.Tech) degree in Electronics and Communication Engineering from Jaypee Institute of Information Technology, Noida. Her research interest is designing of analog circuit.

**Devansh** was born in Delhi, India. He is a 4[th] year student and pursuing a bachelor of technology(B.Tech) degree in Electronics and Communication Engineering from Jaypee Institute of Information Technology, Noida. His research interest is designing of analog circuit.